# Oblivious Transfer Protocol with Verification

Subhash Kak

**Abstract:** Although random sequences can be used to generate probability events, they come with the risk of cheating in an unsupervised situation. In such cases, the oblivious transfer protocol may be used and this paper presents a variation to the DH key-exchange to serve as this protocol. A method to verify the correctness of the procedure, without revealing the random numbers used by the two parties, is also proposed.

1. Introduction

The generation of events of specific probability is essential in many computations and in simulation of physical processes. Of particular interest is the generation of a random sequence that can simulate physical noise and be used for cryptographic and coding purposes. In a random binary (0,1) random sequence, where the bits are independent, the probability of each new bit being 0 (or 1) is ½.

If two parties (Alice and Bob) wish to determine who should play first at a game, they might agree to let Alice play first if she calls the next bit (or the nth future bit) correctly. The problem with this method is that if the algorithm generating the random sequence is known to, say, Alice, she can run it in advance and, therefore, know the next bit in advance. To thwart such a possibility, one would need to place constraints on the nature of the random number generator such as designing it in such a way that it is impossible to emulate it. But that is not a realistic assumption if the generator is an algorithm that is implemented on a computer. If it is easy to generate a pseudo-random sequence, most likely it is cryptographically weak [1]-[7].

Alternatively, one could imagine that a trusted third party has a collection of random number generators. Alice now has to call the *i*th outcome of the *k*th random number generator correctly in order to win the call. If the number of generators is large and the number *i* is derived from some step in a computationally hard number-theoretic problem (such as the number of prime partitions of a large even number), it will become well-nigh impossible for cheating to occur. This is equivalent to the method of puzzles for security [8].

For those who seek mathematical elegance, one might appeal to quantum theory [9]. The outcome of a superposition quantum state, such as $a|0\rangle + b|1\rangle$ is random, with the probability of 0 and 1 being $|a|^2$ and $|b|^2$, respectively. All one needs to do is to start with the state $\frac{1}{\sqrt{2}}(|0\rangle + |1\rangle)$, and measure it along the $|0\rangle$ and $|1\rangle$ bases, and the chosen outcome will have a probability of exactly ½. An example of this are diagonally polarized photons that will be unpredictably received as horizontally or vertically polarized photons along these measurement bases.

This approach via physics is the perfect way to generate random events. But it is not easy to implement [10]-[12]. Due to the Heisenberg's Uncertainty Principle, one cannot generate single quantum states at specified time instants. Indeed, a low-power laser will generate photons with a Poisson distribution [13]. If there are multiple photons with diagonal polarization, the pattern of reduction to the bases states will make it difficult to fix event probabilities. The randomness of collapse is at the basis of quantum cryptography protocols [14],[15]. But due to the difficulty of generating single photon states, quantum cryptography itself uses classical random number generators to guide polarization rotations.

Classical randomness is viewed as an aggregate of countless quantum processes. One could thus have a trusted party look at the thermal noise across a resister at specified future time (so that the bandwidth of the measurement apparatus can be discounted) and check if it is greater or less than the zero threshold. This can serve as an effective method of generating random events. But this requires a trusted third party to supervise the event generation process.

The other method to use is the oblivious transfer (OT) protocol [16],[17] where two parties mutually arrive at the probability event. In the most basic form of OT, the sender sends a message to the receiver with probability ½, while remaining oblivious as to whether or not the receiver obtained the message. Other probabilities can also be likewise generated [18]. These schemes depend on one-way, number-theoretic functions that are at the basis of public key cryptography [19] and they require a choice out of two alternatives to be made at some point in the process.

We assume that the two parties are authenticated to each other and the owner of the secret is honest (the recipient has no reason not being so). To ensure there is no cheating, one could speak in general either of post-communication audit, or supervision of the process by a trusted third party. The audit or verification process should not reveal the random numbers used by the two parties since that could compromise the random number generators used and weaken the security of the process.

We mention parenthetically that randomness was an important notion in ancient societies. The gods were taken to act randomly in a fashion that could not be understood by reasoning. The idea of Vedic ritual [20], Dionysian mysteries, the ecstatic trance of the Oracle of Delphi [21],[22], or shamanic practices of other cultures [23] was to get into a state where one could somehow connect to the time of the gods. The oracle's prophecy was worded ambiguously and what meaning it might convey could not be known to the oracle.

Here we show that an adaptation of the DH key exchange protocol will serve as an OT protocol with verification. We show that the protocol allows Bob to guess Alice's secret with the specified probability. Since the secret belongs to Alice, one can visualize a situation where she cheats so as to reduce Bob's guessing probability. We address this possibility and show how there can be verification of the procedure.

## 2. The Protocol

Alice and Bob together (or a trusted party) choose and publish a large prime $p$ and two integers $u_1$ and $u_2$ of large order modulo $p$. It may thus be assumed that both parties know that $u_1 = k\, u_2$.

> *Step 1.* Alice chooses a random integer $a$, picks one of the two integers $u_1$ and $u_2$ and computes $A = u_i^a \bmod p$, where $i = 1$ or $2$, and sends it to Bob.
>
> *Step 2.* Bob chooses a random integer $b$, picks one of the two integers $u_1$ and $u_2$ and computes $B = u_j^b \bmod p$, where $j = 1$ or $2$, and sends it to Alice.
>
> *Step 3.* Alice takes the received number B and computes $B^a \bmod p = u_j^{ab} \bmod p$ as the key to be used in encrypting a secret file to be sent to Bob.
>
> *Step 4.* Bob takes the received number A and computes $A^b \bmod p = u_i^{ab} \bmod p$ as the key to be used in decrypting a secret file received from Alice.

This protocol is shown in Figure 1 for the special case where Alice and Bob have chosen $u_1$ and $u_2$, respectively. The other cases are where the choice is flipped or where both Alice and Bob choose the same basis.

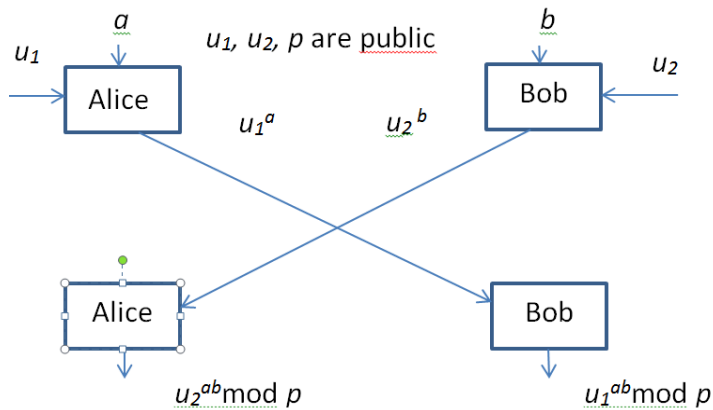

Figure 1. The proposed protocol where Alice and Bob choose different bases

It is assumed that Alice will use the key $u_2^{ab} \bmod p$ to code her secret. She does not know whether Bob possesses this key or $u_1^{ab} \bmod p$. The probability that they choose different bases is ½. Therefore, there is a 0.50 probability that the key generated by Alice and Bob is identical.

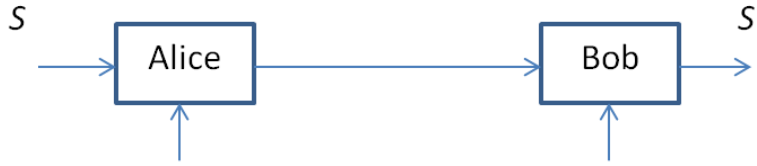

Figure 2. Bob gets the secret, S, if his key is the same as Alice's

If Bob fails to decrypt the secret with his key, he cannot use the knowledge that $u_1 = k\, u_2$, to determine the "correct" key. His incorrect key is related to the correct one through the relationship:

$$u_1^{ab} = u_2^{ab} k^{ab} \bmod p \qquad (1)$$

Bob knows b, k, and $u_1^{ab} \bmod p$, but that is not sufficient to obtain the correct key unless he can solve the discrete logarithm problem.

The eavesdropper also cannot obtain any information about the final key from her observation of the data exchanged by Alice and Bob.

*Generalization.* If in the protocol, there are *m* bases, $u_1, u_2, \ldots, u_m$, rather than just two, as in the example above, the probability that Bob will know the secret is $1/m$.

### 3. Possible cheating by Alice

Alice can cheat by not sending $u_2^{ab} \bmod p$ to Bob over the public channel, but rather $u_2^{fb} \bmod p$, using the exponent *f* to build this fake key. This cheating will be evident if both Alice and Bob choose the same basis, which will happen 50% of the time. The case of cheating thus corresponds to the use of different exponents by the two parties.

To prevent cheating, we add the following steps to the protocol:

> *Step 5.* A random number *r*, publicly declared in advance, is used by Alice to generate $v^n = u_j^{abr} \bmod p$ (n=abr). In the example of Figure 1, $v^n = u_2^{abr} \bmod p$. The number $v^n$ is sent to Bob.

> *Step 6.* Bob uses the verification sequence $G(n) = v^n + w^n \bmod p$ to establish that there has been no cheating.

If $v = w$, $G(n) = 0$. When $v \neq w$, $G(n) = \alpha\, G(n-1) + \beta\, G(n-2) \bmod p$, where α and β are constants that are easily found. The verification sequence G(n) is described in the next section.

If Alice were to cheat by using $u_2^{fb} \bmod p$ as the key, but sends the correct $u_2^n \bmod p$, she will be exposed in case Bob has chosen $u_2$ and finds G(n) =0, while remaining unable to decrypt the secret.

4. **The verification sequence** $G(n)$

Consider the sequence $G(n) = v^n + w^n \bmod p$. In general we can write

$$v^k = \alpha_k v + \beta_k \bmod p$$
$$w^k = \alpha_k w + \beta_k \bmod p \qquad (2)$$

**Theorem 1**. $G(n) = \alpha_k G(n-k+1) + \beta_k G(n-k) \bmod p \qquad (3)$

*Proof.* $G(n) = (v^n + w^n) \bmod p$
$$= (v^{n-k} v^k + w^{n-k} w^k)$$
$$= v^{n-k}(\alpha_k v + \beta_k) + w^{n-k}(\alpha_k w + \beta_k)$$
$$= \alpha_k(v^{n-k+1} + w^{n-k+1}) + \beta_k(v^{n-k} + w^{n-k})$$
$$= \alpha_k G(n-k+1) + \beta_k G(n-k) \bmod p$$

When $k = 2$,

$$G(n) = \alpha\, G(n-1) + \beta\, G(n-2) \bmod p \qquad (4)$$

which means that the sum of successive powers of *v* and *w* suffices to establish that they have been computed to the same exponent. All that is required to find the values of α and β is the solution to equation (2) for $k = 2$. No knowledge of the actual value of n is needed while computing equation (4).

*Example 1*. Let $k=2$, $v=3$, and $w=7 \bmod 19$. To find *α* and *β*, we solve the equations:

$$3^2 = 9 = \alpha 3 + \beta \bmod 19$$
$$7^2 = 11 = \alpha 7 + \beta \bmod 19$$

We find that *α*=10 and *β*=17.

The series $G(n) = 3^n + 7^n \bmod 19$, for n = 0, 1, 2, 3,… is as follows:

2, 10, 1, 9, 12, 7, 8, …

for which each *n*th element is 10 G(n-1)+13 G(n-2) mod 19. For example, the value 9 is 10×1+17×10 mod 19.

*Example 2*. Let $k=2$, $v=3$, and $w=5 \bmod 17$. To find *α* and *β*, we solve the equations:

$$3^2 = 9 = \alpha 3 + \beta \bmod 17$$
$$5^2 = 8 = \alpha 5 + \beta \bmod 17$$

We find that $\alpha=8$ and $\beta=2$.

The series $G(n) = 3^n + 5^n \mod 17$, for n=0,1,2,3… is as follows:

    2, 8, 0, 16, 9, 2, 0, 4, 15, 9, …

for which each nth element is 8 G(n-1)+2 G(n-2) mod 17.

Theorem 1 may be extended to modulo *m*, if *u* and *v* are relative prime to *m*.

If the exponents in equation (2) are not the same then the result of Theorem 1 will not be valid.

Since *v* and *w* are known, three consecutive G(n) values can be computed by successive multiplication with the appropriate bases and it checked if the successive numbers have the relationship of equation (3).

## 5. Conclusions

This paper reviews the problem of generation of random events using classical and quantum techniques. It then presents a variation of the DH key exchange protocol to serve as an oblivious transfer protocol that can easily generate a probability event of 1/m, where m is 2 or higher integer. A verification procedure is presented that can catch attempts by Alice at cheating.

*Acknowledgement*. This research was supported in part by research grant #1117068 from the National Science Foundation.